\documentclass[prb,preprint,aps]{revtex4}
\usepackage{amsmath}
\usepackage{graphicx}
\usepackage[dvips]{color}
\usepackage{braket}

\begin{document}
\title{Quantum metamaterials in the microwave and optical ranges}
%
%
\author{A.M. Zagoskin$^{1,2}$}
\author{Didier Felbacq$^{3}$}
\author{Emmanuel Rousseau$^{3}$}

\affiliation{(1) Department of Physics, Loughborough University,
Loughborough LE11 3TU, United Kingdom}
\affiliation{(2) Theoretical Physics and Quantum Technologies Department,
Moscow Institute for Steel and Alloys, 119049 Moscow, Russia}
\affiliation{(3) University of Montpellier, Laboratory Charles Coulomb UMR CNRS-UM 5221, Place Bataillon, B\^at. 21 CC074, 34095 Montpellier Cedex 05, France}

\date{\today}

\begin{abstract}
Quantum metamaterials generalize the concept of metamaterials (artificial optical media) to the case when their optical properties are determined by the interplay of quantum effects in the constituent 'artificial atoms' with the electromagnetic field modes in the system. The theoretical investigation of these structures demonstrated that a number of new effects (such as quantum birefringence, strongly nonclassical states of light, etc)  are to be expected, prompting the efforts on their fabrication and experimental investigation. Here we provide a summary of the principal features of quantum metamaterials and review the current state of research in this quickly developing field, which bridges quantum optics, quantum condensed matter theory and quantum information processing.
\end{abstract}
 \maketitle

\tableofcontents

\section{Introduction}

The turn of the century saw two remarkable developments in physics. First, several types of scalable solid state quantum bits were developed, which demonstrated controlled quantum coherence in artificial mesoscopic structures\cite{Nakamura1999,Mooij1999,Friedman2000,Hayashi2003} and eventually led to the development of structures, which contain hundreds of qubits and show signatures of global quantum coherence (see \cite{Zagoskin2013a,Albash2015} and references therein). In parallel, it was realized that the interaction of superconducting qubits with quantized electromagnetic field modes reproduces, in the microwave range, a plethora of effects known from quantum optics (in particular, cavity QED) with qubits playing the role of atoms ('circuit QED', \cite{Blais2004b,Blais2004a,Blais2007}). 
 Second, since John Pendry\cite{Pendry2000} extended the results by Victor Veselago\cite{Veselago1968}, there was an explosion of research of classical metamaterials resulting in, e.g., cloaking devices in microwave and optical range \cite{Greenleaf2009,Alitalo2013,Fleury2015}. The logical outcome of this parallel development was to ask, what would be the optical properties of a "quantum metamaterial" - an artificial optical medium, where the quantum coherence of its unit elements plays an essential role? 

As could be expected, this question was arrived at from the opposite directions, and the term "quantum metamaterial" was coined independently and in somewhat different contexts. In refs.\cite{Plumridge2007,Plumridge2008,Plumridge2008a} it was applied to the plasmonic properties of a stack of
2D layers, each of them thin enough for the motion of electrons in the normal
direction to be completely quantized. Therefore "the wavelike nature of matter"
had to be taken into account at a single-electron level, but the question of
quantum coherence in the system as a whole did not arise. In refs.\cite{Rakhmanov2008,Zagoskin2009} the starting point was the explicit
requirement that the system of artificial atoms (qubits) maintained quantum coherence on the time scale of the electromagnetic pulse propagation
across it, in the expectation that the coherent quantum dynamics
of qubits interacting with the electromagnetic field governed the "optical" properties of the metamaterial. 

Currently the term "quantum metamaterial" is being used in both senses (see, e.g., \cite{Zheludev2010,Quach2011,Felbacq2012,Zagoskin2012,Savinov2012,Zheludev2012}). We will follow the more restrictive usage and call quantum
metamaterials (in the narrow sense) such artificial optical (in the broad sense) media that\cite{Zagoskin2012}
(i) are comprised of quantum coherent unit elements with desired (engineered)
parameters;
(ii) quantum states of (at least some of) these elements can be directly controlled;
and
(iii) can maintain global coherence for the duration of
time, exceeding the traversal time of the relevant electromagnetic signal.
The totality of (i)-(iii) (in short: controlled macroscopic quantum coherence)
that makes a quantum metamaterial a qualitatively different system, with a
number of unusual properties and applications.

A conventional metamaterial can be described by effective macroscopic parameters,
such as its refractive index. (The requirement that the size of a unit cell of the system be much less - in
practice at least twice less - than the wavelength of the relevant electromagnetic
signal, is implied in its definition as an optical medium, and is inherited by quantum metamaterials.)  From the microscopic point of view,
these parameters are functions of the appropriately averaged quantum states
of individual building blocks. In a quantum metamaterial, these states can be
directly controlled and maintain phase coherence on the relevant spatial and
temporal scale. 

The full treatment of such a system should start from quantum description of both the electromagnetic field and the "atoms". In case when their role is played by qubits (that is, two-level quantum systems), a general enough Hamiltonian of a quantum metamaterial is given by\cite{Quach2011,Mukhin2013}
\begin{equation}
\hat{H} = \sum_j \hbar\omega_j b_j^+b_j - \frac{1}{2} \sum_k \left(\epsilon_k\sigma_z^{(k)} + \Delta_k \sigma_x^{(k)}\right) + i \sum_{jk} \xi_{jk}\sigma_z^{(k)}\left(b_j-b_j^+\right),
\label{eq:1}
\end{equation}
where the first term describes unperturbed photon modes, the second the qubit degrees of freedom, and the third their interaction. The direct control over (at least some) qubits is realized through the qubit Hamiltonian parameters (the bias $\epsilon$ or, sometimes, the tunneling matrix element $\Delta$).

Forming  photon wave packets with characteristic size $\Lambda \gg a$ (where $a$ is the unit cell size) and averaging over the quantum states of qubits on the scale of $\Lambda$, one should eventually arrive at the effective equation of motion for the "$\Lambda$-smooth" density matrix. It will describe the state of both the electromagnetic field and the quantum metamaterial, characterized by a nondiagonal, nonlocal, state- and position-dependent "refractive index" matrix. 

Following through with this program involves significant technical difficulties, and the task is not brought to conclusion yet. Nevertheless certain key effects in quantum metamaterials can be investigated at a more elementary level using an approximate wave function (e.g., \cite{Zhou2008,Quach2009,Mukhin2013,Biondi2014}), or treating the electromagnetic field classically\cite{Rakhmanov2008,Zagoskin2009,Zagoskin2011,Asai2015}. The latter is like the standard quasi-classical treatment of the atom-light interaction\cite{Blokhintsev1964}, but is more conveniently done using the lumped-elements description (see, e.g., \cite{Zagoskin2011}, Section 2.3).  The state of a system of $M$ nodes connected by capacitors, inductors and, if necessary, Josephson junctions, is described by a Lagrangian ${\cal L}(\{\Phi,\dot{\Phi}\})$.
Here the "node fluxes" $\Phi_j(t) = c\int^t dt'\; V_j(t')$ are related to the node voltages $V_j(t)$ and completely describe the classical electromagnetic field degrees of freedom (the current-voltage distribution) in the system. The lumped-elements description is appropriate, since we are interested in signals with a wavelength much larger than the dimension of a unit element of the circuit (the condition of its serving as a metamaterial).

Qubits are introduced in this scheme through their own Hamiltonians and coupling terms (e.g., flux qubits can be coupled inductively - through the magnetic flux penetrating their loop, - or galvanically - through sharing a current-carrying conductor with the circuit), which are usually cast in the form
\begin{equation}
\hat{H}_q =  -\frac{1}{2} \left[\epsilon_q \sigma_q^z + \Delta_q \sigma_q^x\right] + J_q(\Phi,\dot{\Phi})\sigma_q^z.
\label{eq:2}
\end{equation}
The "quantum Routhian", 
\begin{equation}
\hat{\cal R}(\{\sigma\},\{\Phi,\dot{\Phi}\}) = \sum_q \hat{H}_q - {\cal L}(\{\Phi,\dot{\Phi}\}),
\label{eq:3}
\end{equation}
plays the role of the Hamiltonian for qubits (in Heisenberg representation), and its expectation value ${\cal R} =\langle \hat{\cal R} \rangle$ produces the equations of motion for the classical field variables $\Phi$, where we have included the dissipative function $\cal Q$ to take into account resistive losses in the circuit (see Refs.\cite{Zagoskin2011}, 2.3.3, and \cite{Zagoskin2014}):
\begin{equation}
\frac{d}{dt} \frac{\partial{\cal R}}{\partial \dot{\Phi}_j} - \frac{\partial{\cal R}}{\partial \Phi_j} = \frac{\partial{\cal Q}}{\partial \dot{\Phi}_j}.
\label{eq:4}
\end{equation}
If one approximates the quantum state of the qubit subsistem by a factorized function, one can take the continuum limit and, switching to the Schr\"{o}dinger representation, obtain coupled systems of equations for the field variable $\Phi({\bf r},t)$ and the qubit two-component "macroscopic wave function" $\hat{\Psi}({\bf r},t)$. At any convenient stage the electromagnetic modes can be canonically quantized via $\Phi \to \hat{Q} \sim (b+b^{\dag})$, $\dot{\Phi} \to \hat{P} \sim i(b-b^{\dag})$, leading back to a certain approximation of an approach based directly on Eq.(\ref{eq:1}).

\section{Superconducting quantum metamaterials}

\begin{figure}
\includegraphics[width=\columnwidth]{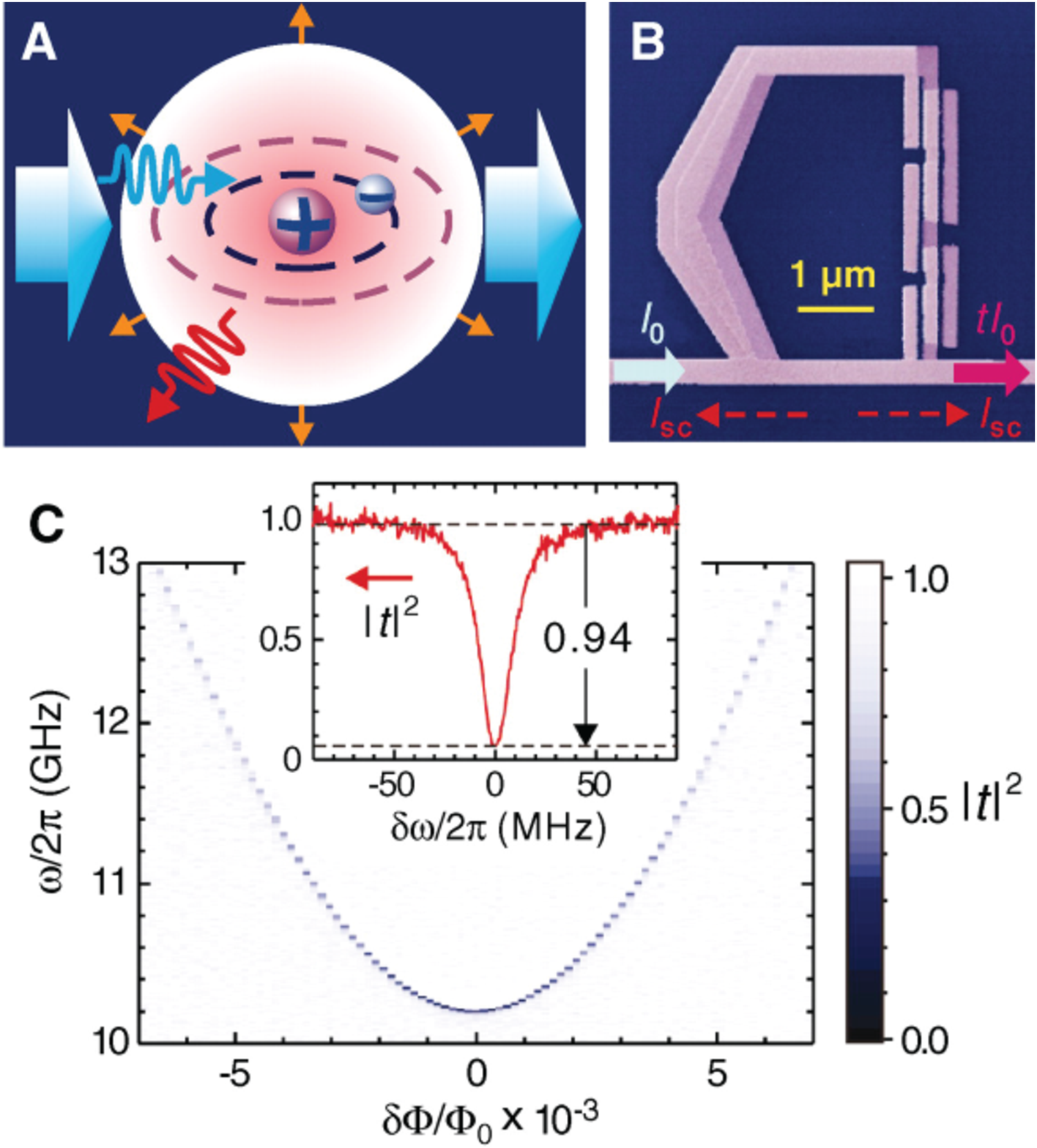}
\includegraphics[width=\columnwidth]{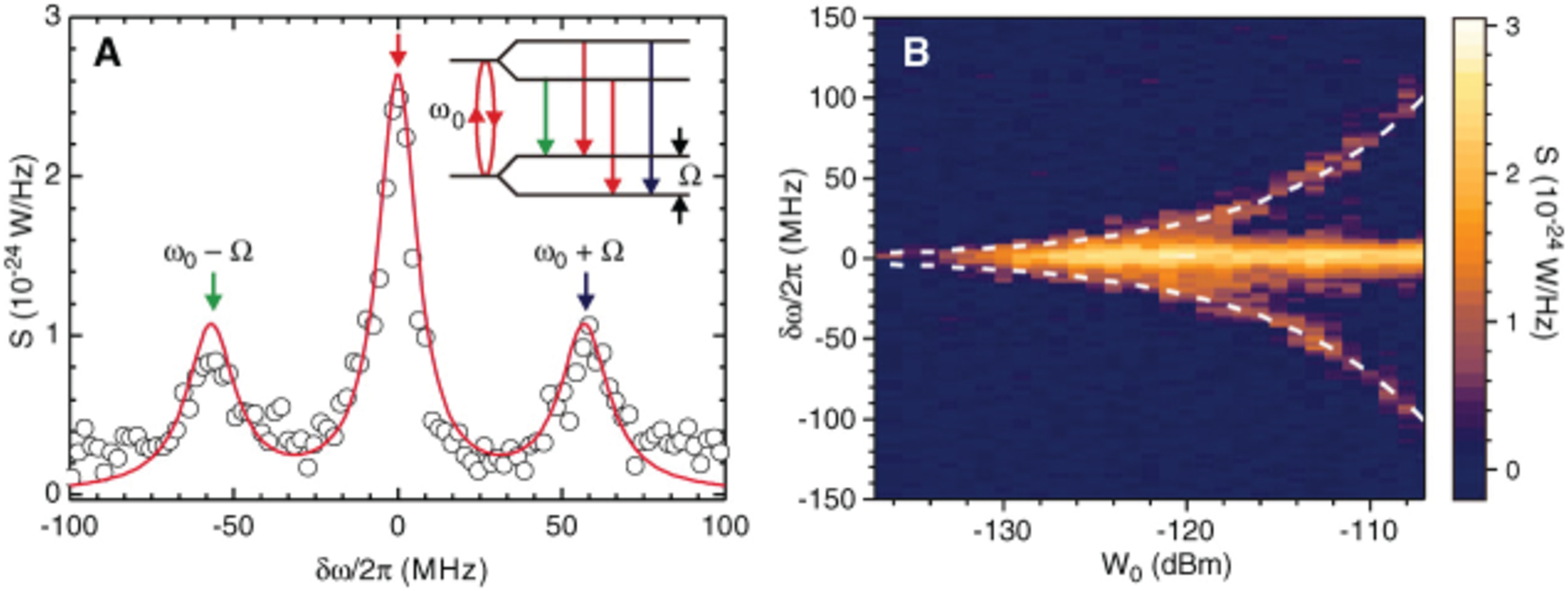}
\caption{(a) A "proto-metamaterial": a single artificial atom (flux qubit) in a transmission line. (b) Reflection coefficient of a microwave signal in the line measured (top) and calculated (bottom) for the structure of Fig.(a). After Ref.\onlinecite{Astafiev2010} }. 
\label{fig:1}
\end{figure}

The above scheme is most often applied - though not limited - to the case of superconducting metamaterials based on various types of superconducting qubits (see, e.g., Ref.\onlinecite{Zagoskin2011}, Ch.2). The simplest case is given by the experimental setup of Ref.\cite{Astafiev2010} (Fig.\ref{fig:1}a). There a single artificial atom (a flux qubit) is placed in a transmission line, and transmission and reflection coefficients for the microwave signal are measured. Here the equations (\ref{eq:4}) for the field in the continuum limit will yield free telegraph equations for the voltage and current everywhere except the point $x=0$, where the qubit is situated:
\begin{eqnarray}
\frac{\partial V(x,t)}{\partial x} = \frac{\tilde{L}}{c^2} \frac{\partial I(x,t)}{\partial t}; \\
\frac{\partial I(x,t)}{\partial x} = \tilde{C} \frac{\partial V(x,t)}{\partial t},
\label{eq:4a}
\end{eqnarray}
where $\tilde{L}, \tilde{C}$ are the inductance and capacitance per unit length of the transmission line. Of course, for such a simple structure these equations can be written down directly. The influence of the flux qubit, which is coupled to the transmission line through the effective mutual inductance $M$ (taking into account both magnetic and kinetic inductance) is through the matching conditions at $x=0$,
\begin{eqnarray}
V(+0,t) = V(-0,t) - \frac{M}{c}\frac{\partial \langle \hat{I}_q (t) \rangle}{\partial x}; \\
I(+0,t) = I(-0,t).
\label{eq:5}
\end{eqnarray}
The qubit current operator $\hat{I}_q = I_p\sigma^z$ is governed by the qubit Hamiltonian (\ref{eq:2}) with the coupling term $J_q = I(0,t)M/c^2$. An explicit solution for the reflection/transmission amplitudes was found to be in a  very good agreement with the experimental data in Ref.\onlinecite{Astafiev2010} (Fig.\ref{fig:1}b).

For a structure, where the role of the "artificial atom" between 1D transmission lines is played by a single qubit surrounded by an array of $N$ coupled photonic cavities \cite{Biondi2014} the calculations in the one-excitation approximation (with electromagnetic modes treated quantum mechanically) show that in such a structure arise long-living quasi-bound states of photons and the qubit, manifested as ultra-narrow resonances in the transmission coefficient. 

\begin{figure}%
\includegraphics[width=\columnwidth]{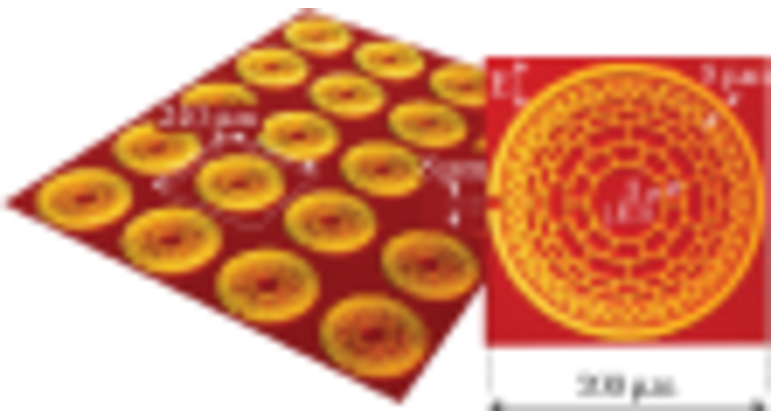}
\includegraphics[width=\columnwidth]{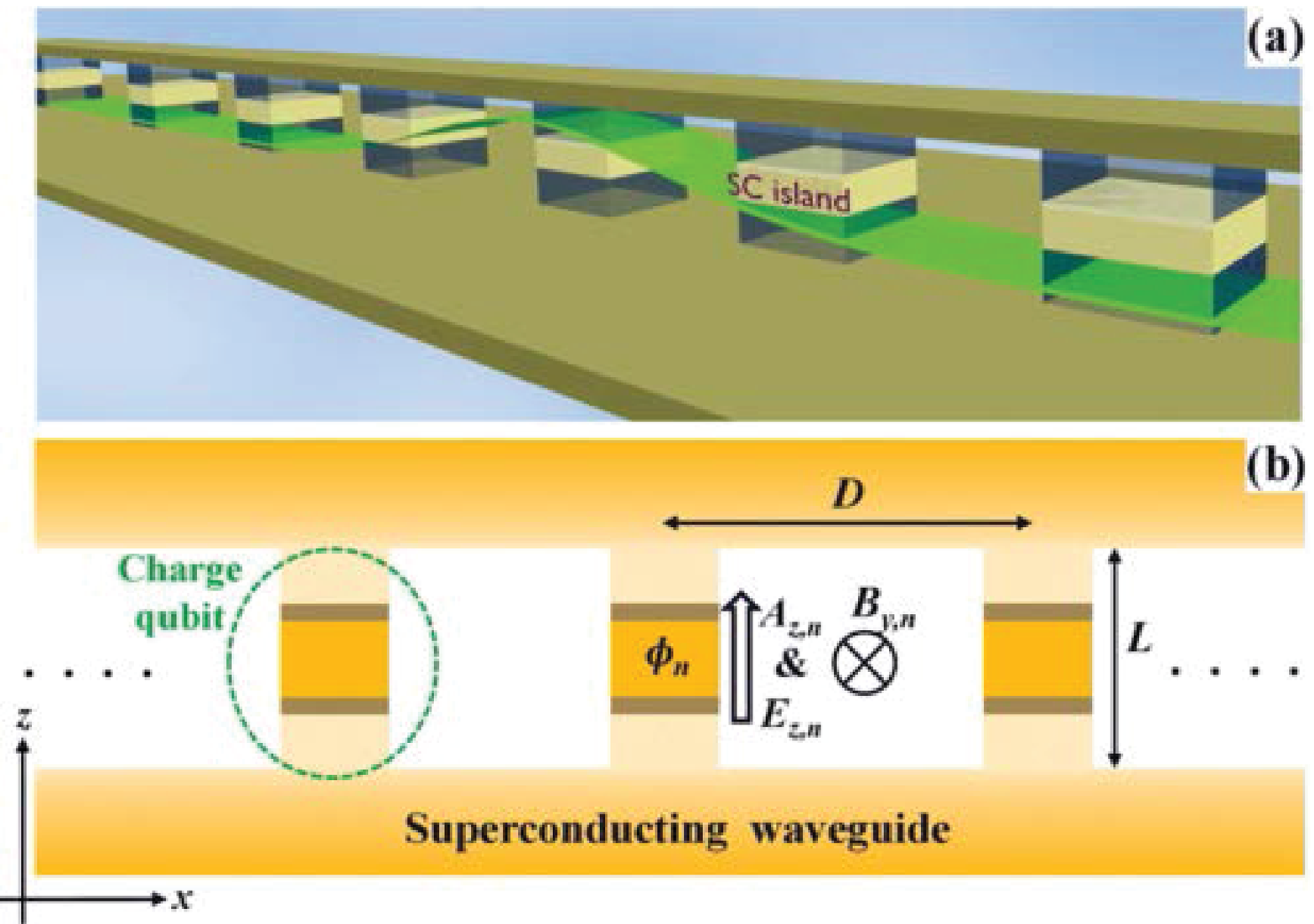}
\caption{(a) A 2D superconducting quantum metamaterial \cite{Savinov2012}. \textcolor{red}{Fig.3 of \cite{Savinov2012}} (b) A realization of a 1D superconducting quantum metamaterial: charge qubits in a transmission line \cite{Asai2015}.} 
\label{fig:2}%
\end{figure}

\begin{figure}
\includegraphics[height=5cm]{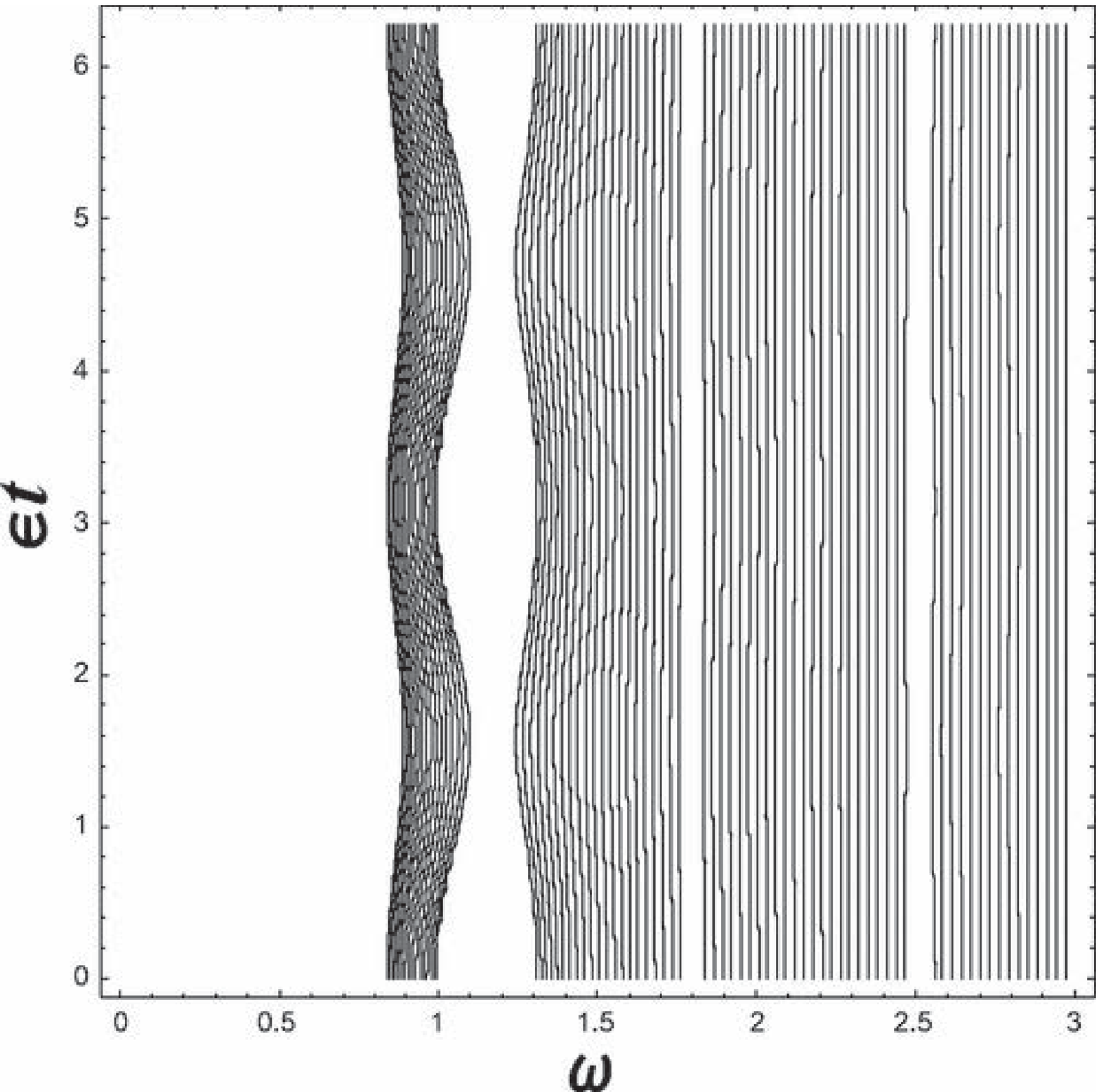}
\includegraphics[height=5cm]{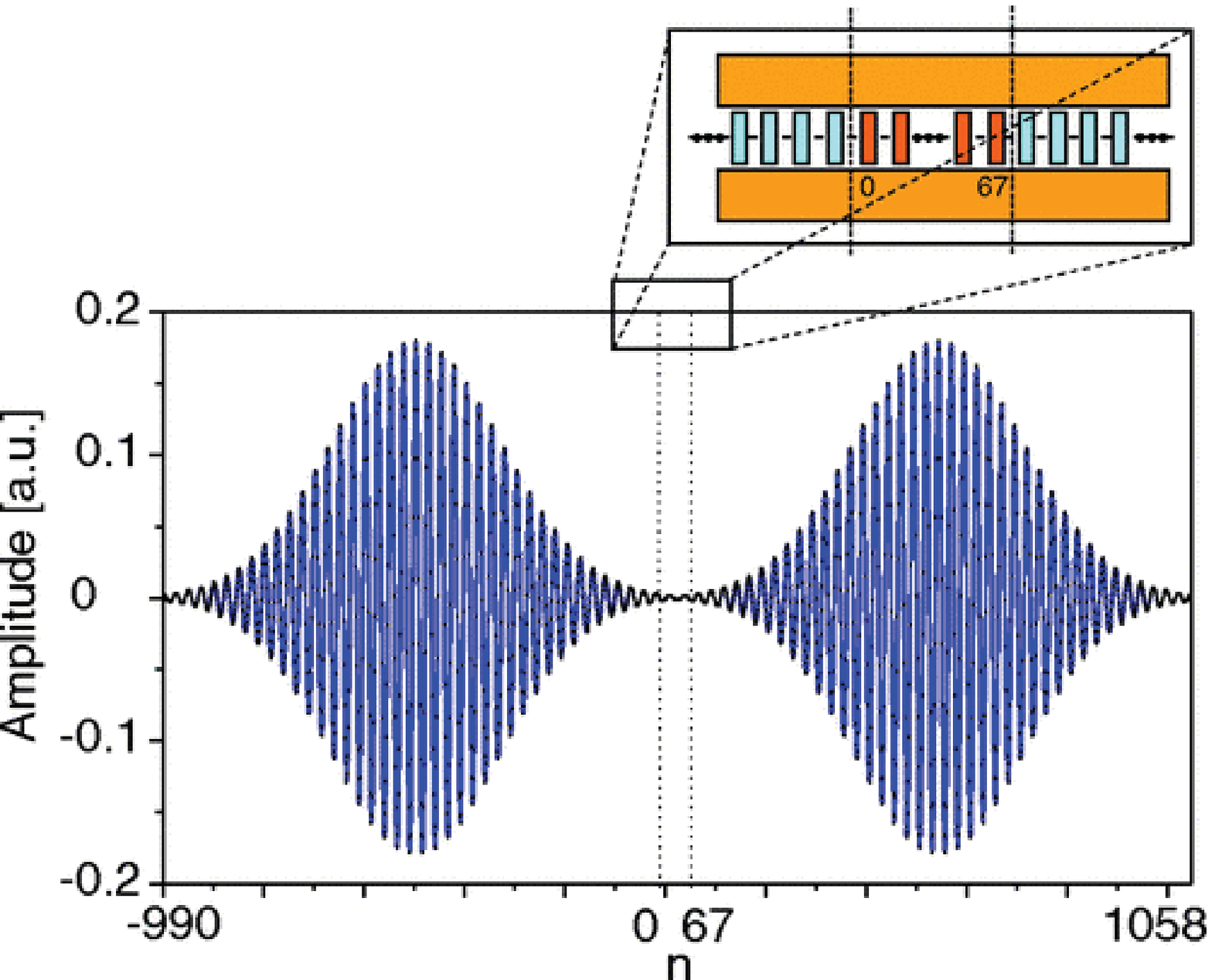}
\includegraphics[height=5cm]{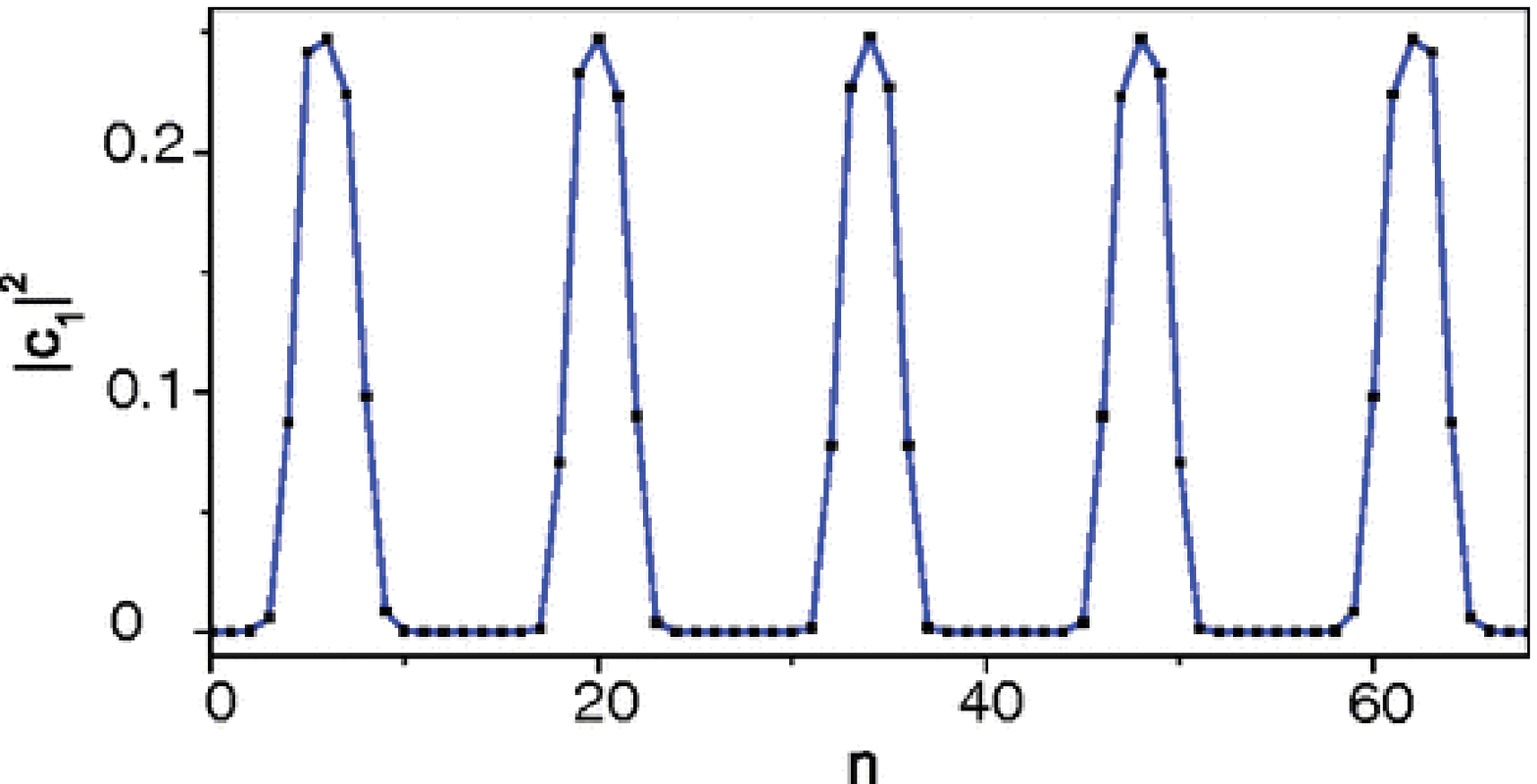}
\includegraphics[height=5cm]{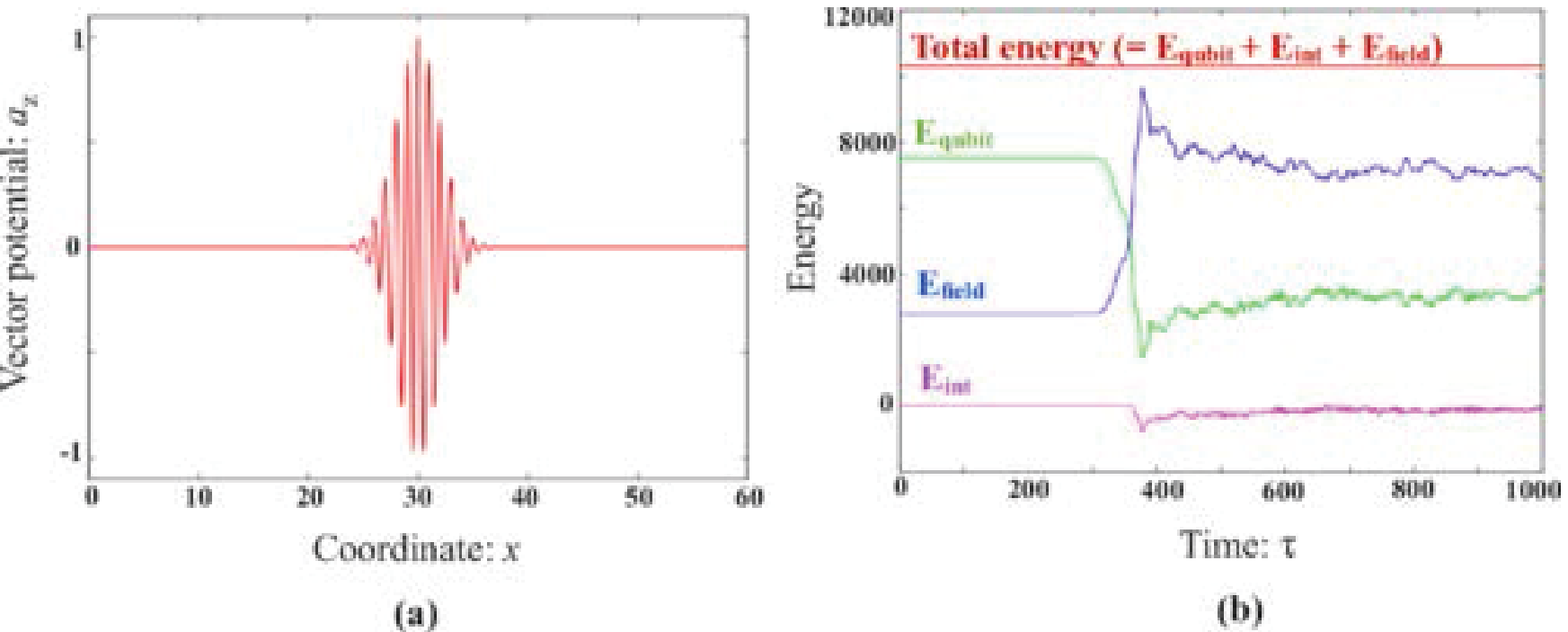}
\includegraphics[height=5cm]{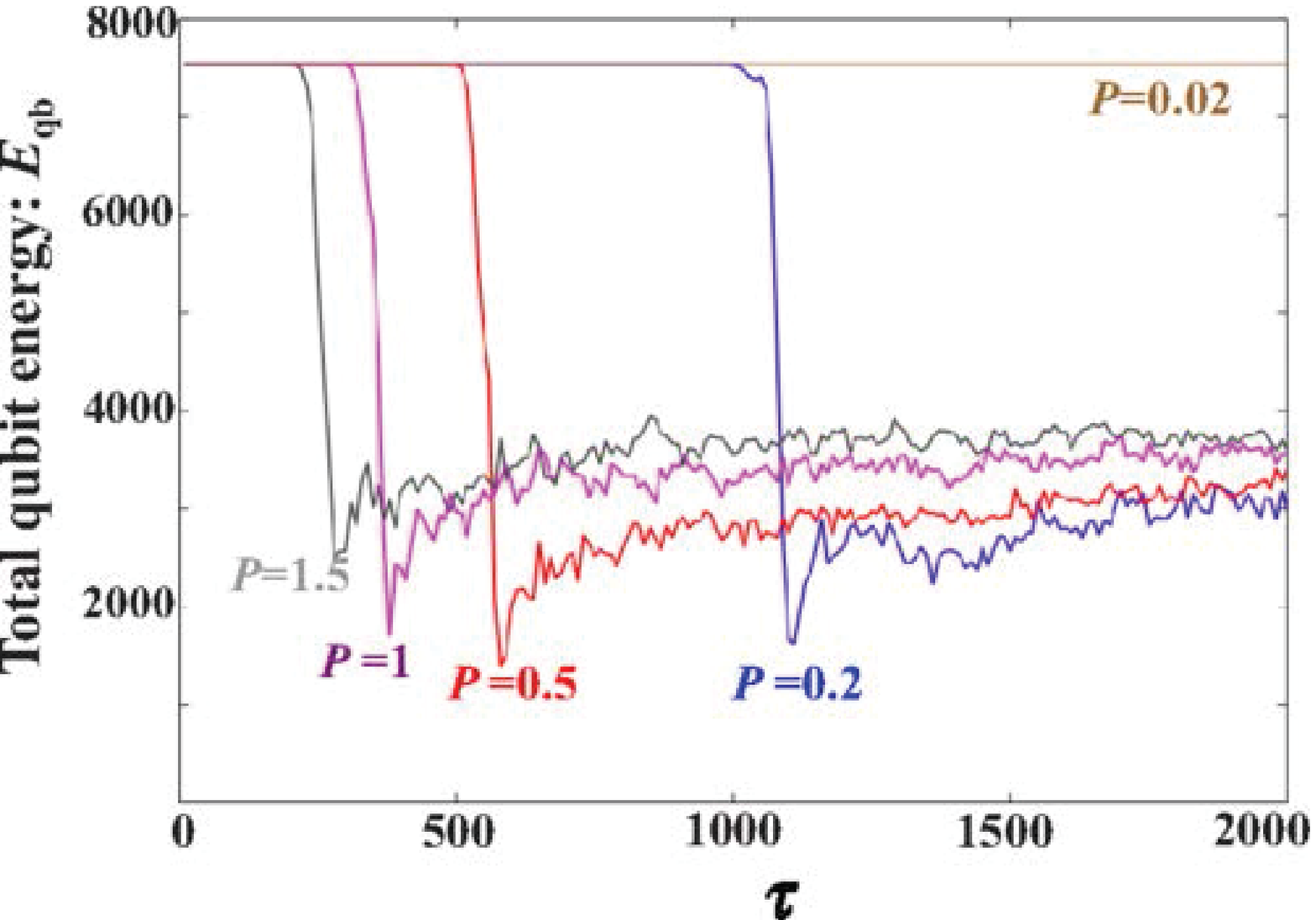}
\caption{(a) Breathing photonic crystal \cite{Rakhmanov2008}. 
(b) Creating a photonic crystal without a direct control of a QMM \cite{Shvetsov2013}.
(c) Lasing in a QMM \cite{Asai2015}. }%
\label{fig:3}
\end{figure}

Going from these "proto-metamaterials" to QMMs containing many artificial atoms, we return to the classical treatment of the electromagnetic field. Though 2D and 3D versions of superconducting QMMs are feasible \cite{Savinov2012,Zagoskin2012} (Fig. \ref{fig:2}a), they do not operate yet in a quantum coherent regime. Most of the research at the moment concentrates on the 1D case, which already promises interesting results. In theoretical papers \cite{Rakhmanov2008,Shvetsov2013,Asai2015} is considered a QMM formed by a set of superconducting charge qubits placed in the transmission line (Fig.\ref{fig:2}b). The equations of motion for the field and qubits were solved, using a factorized approximation of the quantum state vector of the qubit subsystem. For the realistic choice of qubit and transmission line parameters, the figures of merit $\beta = \sqrt{E_{EM}/
\Delta} \sim 30$ and $\nu = \hbar\max(\Gamma_{qb},\Gamma_{TL})/\Delta \sim 10^{-3}$  (Ref.\onlinecite{Rakhmanov2008}) ensure that the continuum approximation is justified, and that in the first approximation the effects of decoherence can be neglected. Here $E_{EM}$ is the electromagnetic field energy per unit cell, $\Delta$(Eq.(\ref{eq:1})) gives the qubit energy scale, $\beta$ is the dimensionless signal velocity in the transmission line (units cells per $\hbar/E_J$), and $\Gamma_{qb}, \Gamma_{TL}$ are the decoherence rates in a qubit and in the transmission line respectively. Some of the results are shown in Fig.\ref{fig:3}. 

Dimensionless equations of motion for the field vector-potential in the lowest order in field-qubits interaction yield the wave equation
\begin{equation}
\ddot{\alpha}(\xi,\tau)  - \beta^2 \frac{\partial^2\alpha(\xi,\tau)}{\partial \xi^2} + V(\xi,\tau)\alpha(\xi,\tau) = 0,
\label{eq:6}
\end{equation}
where $V(\xi,\tau) \propto \langle\hat{\Psi}(\xi,\tau)|\cos\phi|\hat{\Psi}(\xi,\tau)\rangle$. The tunneling matrix element of the charge qubit Hamiltonian ($\phi$ being the qubit superconducting phase) is determined by the qubit state $|\hat{\Psi}(\xi,\tau)\rangle$ at the given point. The dispersion law is thus directly dependent on the QMM quantum state, as expected. For example, if the qubit state is a periodic function of the coordinate, $\hat{\Psi}(x+\Lambda)=\hat{\Psi}(x)$, the QMM behaves as a photonic crystal, with gaps opening in the electromagnetic spectrum\cite{Rakhmanov2008,Shvetsov2013}, which can be manipulated by controlling the quantum state of qubits. For example, if qubits are placed in a spatially periodic superposition of their eigenstates,
\begin{equation}
|\hat{\Psi}(x,t)\rangle = A(x)|g\rangle + B(x)|e\rangle e^{-i\omega t},
\label{eq:7}
\end{equation}

where $A(x), B(x)$ are periodic functions and $\hbar\omega$ is the energy splitting between the ground ($|g\rangle$) and excited ($|e\rangle$) state of the qubit. Quantum beats between these states will produce a "breathing" photonic band structure (Fig.\ref{fig:3}a); external control of qubit states allows to trap a portion of radiation in a pocket of such a structure and move it across the QMM at a desired speed \cite{Rakhmanov2008}. In the absence of direct control over individual qubit states (apart from their initialization in the ground state) it is still possible to create a photonic crystal structure \cite{Shvetsov2013} by sending into the QMM specially shaped "priming" electromagnetic pulses from the opposite directions (Fig. \ref{fig:3}b). Exiting the QMM, they leave behind a spatially periodic pattern of the probability of finding a qubit in the ground (excited) state.

Solving the coupled equations for the classical field and qubits numerically (still in the approximation of factorized qubit state) allows to investigate lasing in a QMM \cite{Asai2015}. If the qubits are initialized in the excited  state (e.g. by sending a priming pulse through the QMM), an initial pulse triggers a coherent transition of energy from qubits to the electomagnetic field (Fig. \ref{fig:3}c. Remarkably, not only the process has a precipitous character, but its onset starts the sooner the greater the amplitude of the triggering pulse: $\sqrt{{\rm field}\:{\rm amplitude}} \times \tau_{\rm onset} \approx {\rm const}$.

In equilibrium a fully quantum treatment of a superconducting QMM becomes possible, which allows the investigation of phase transitions in the photon system \cite{Mukhin2013}. The chosen model of a QMM (a series of RF SQUIDS coupled to the transmission line and considered in two-level approximation) lead to a generic Hamiltonian (\ref{eq:1}), with only parameters being model dependent. Using the instanton approach, the effective action of the photon subsystem was obtained as a function of the photon field momentum $P$ (in imaginary time). 

In case when $P_0$ is independent on the imaginary time $\tau$, the photon system may undergo a second order classical phase transition: above a critical temperature $T^{\star}$ the momentum $P=0$, while below it the system can choose between two values $\pm P_0$. The transition details depend on the level of disorder in the qubit chain, which is modeled here by the random distribution of tunneling matrix elements $\Delta$. In the case of a low disorder, $\Delta_k \approx \Delta_0$, the transition occurs at the critical temperature
\begin{equation}
T^{\star}_n = \frac{m\eta^2N}{k_B},
\label{eq:8}
\end{equation} 
where $m$ is the transmission line inductance per unit cell, $\eta$ parametrizes the qubit-field interaction, and $N$ is the total number of qubits in the QMM. 
This phase transition occurs only if $\Delta_0 < k_BT^{\star}_n$. In the case of strong disorder with $\Delta_k$ distributed from zero to some $\Delta_0$, the transition temperature 
\begin{equation}
T^{\star}_w = \frac{\Delta_0 \exp\left[-\Delta_0/m\eta^2N\right]}{k_B},
\label{eq:9}
\end{equation} 
and the transition occurs only if $\Delta_0 > m\eta^2N$. As the authors of Ref.\onlinecite{Mukhin2013} note, the first case is similar to the metal-ferromagnet phase transition, while the second one is reminiscent of the normal metal-superconductor or Peierls metal-insulator transition. In either case a coherent state of the photon field emerges, with nonzero value of the order parameter $P_0$, which in the case of low disorder is proportional to the number of qubits in the system, $N$.

\begin{figure}%
\includegraphics[width=\columnwidth]{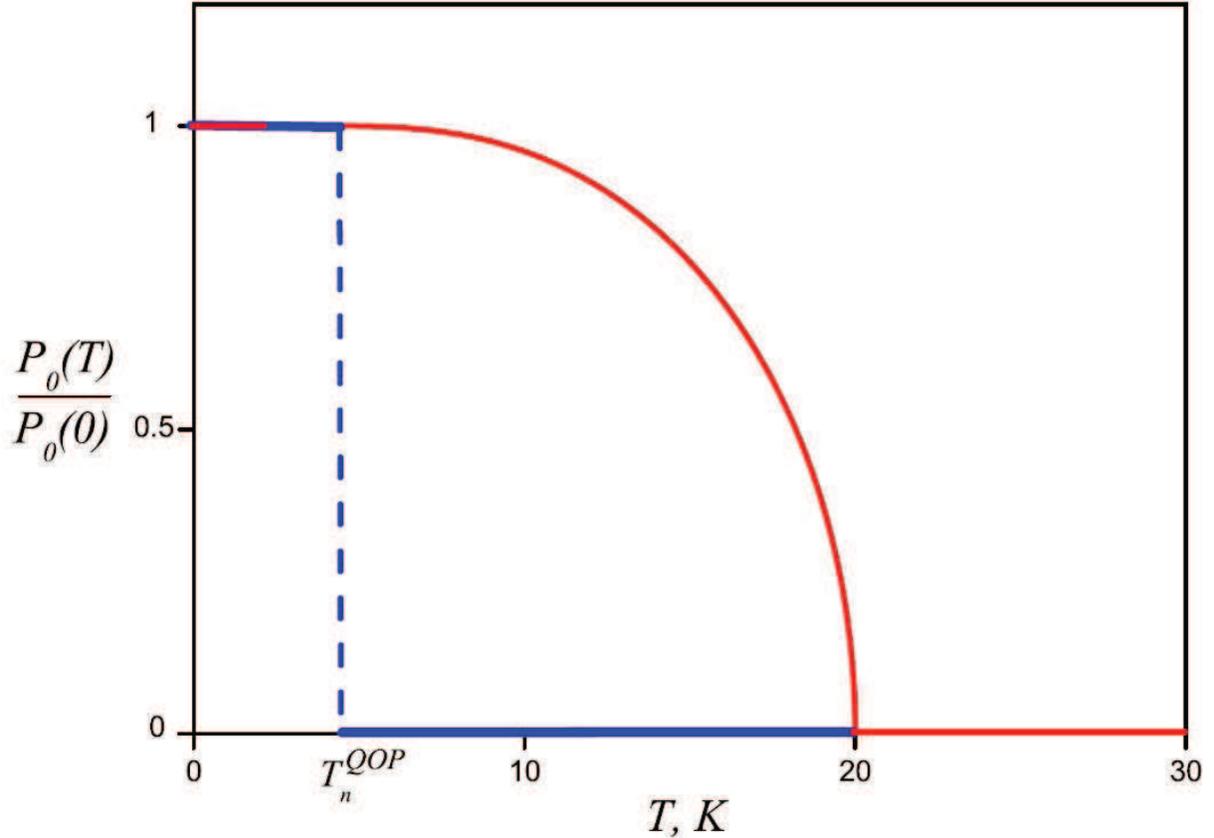}
\caption{Classical and quantum phase transition in the state of photons in a 1D superconducting quantum metamaterial\cite{Mukhin2013}. 
The photon order parameter $P_0(\tau)$ as a function of imaginary time $\tau$ is either constant (classical phase transition, red line, $P_0 \propto N$) or proportional to $\sin [P_0(T) \tau]$ (quantum phase transiion, blue line, $P \propto N^{1/3}$). The low disorder case is shown: $\Delta_k \approx \Delta_0 = 4$K; $T^{\star}_n = 20$K.}%
\label{fig:4}%
\end{figure}

It turns out that there is the possibility of a quantum transition as well, into a state representing a superposition of semiclassical states $\pm P_0$. Then the order parameter becomes a periodic function of $\tau$. In the case of low disorder and strong field-qubit coupling it occurs below the temperature (Fig.\ref{fig:4})
\begin{equation}
T^{QOP}_n \propto \left[\frac{\Delta_0^2m\eta^2N}{\pi^2}\right]^{1/3} \frac{1}{k_B},
\label{eq:10}
\end{equation}
while in the opposite case it is given by
\begin{equation}
T^{QOP}_n = \frac{\Delta_0}{k_B\pi} \exp\left[-\Delta_0/m\eta^2N\right].
\label{eq:11}
\end{equation}

The estimates of Ref.\onlinecite{Mukhin2013} for the transition temperatures $T^{\star} \sim 0.1 \dots 50$ K let us hope for direct observation of photon phase transitions in a superconducting QMM, when the number of qubits, the homogeneity of their parameters and their coherence times are improved as compared to the existing QMM prototype \cite{Macha2014,Ustinov2015}. The prototype (Fig.5a) consists of 20 flux qubits placed in a coplanar waveguide resonator. The inductive coupling of qubits to the resonator and each other was comparable, but strong decoherence (decoherence time of the order of a few nanoseconds) effectively suppressed qubit-qubit coupling. Nevertheless the transmission measurements in the resonant regime showed the formation of three ensembles of interacting qubits (two of four qubits each and one of eight qubits). This interaction through the electromagnetic modes is the key element of the operation of a QMM. In order to improve the operation of a QMM it is suggested to increase the qubit-qubit coupling, in order to counteract the effects of qubit parameter dispersion.

\begin{figure}%
\includegraphics[height=7cm]{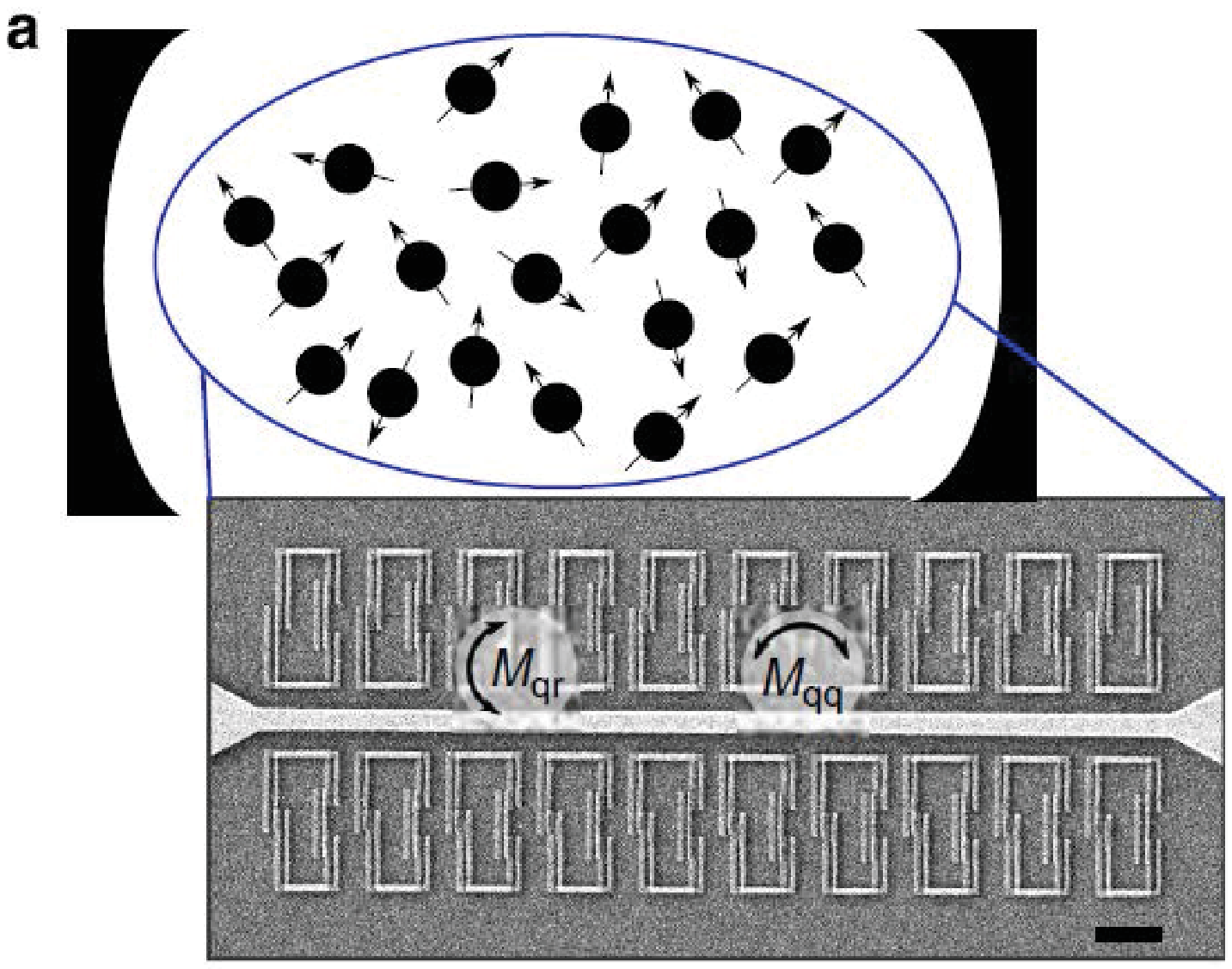}
\includegraphics[height=7cm]{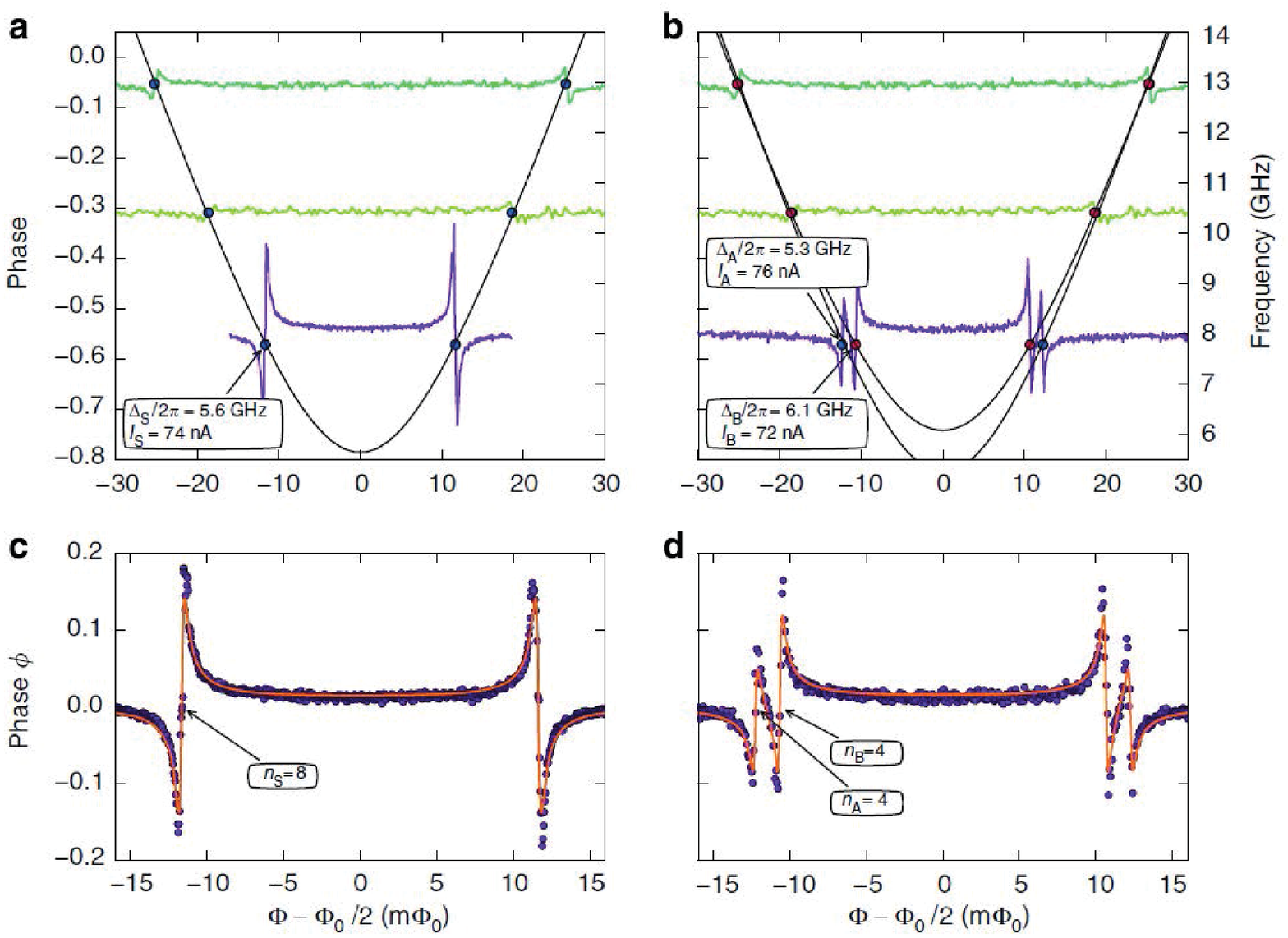}
\caption{Quantum metamaterial prototype \cite{Macha2014}. 
(a) Twenty flux qubits placed in a resonator. The signal wavelength is of the order of the resonator length (23 mm) and greatly exceeds the qubit size (approx. $2 \times 6$ $\mu$m). The ensemble average level spacing and persistent current in the qubits are 5.6 GHz and 74 nA respectively. (b) The QMM in the resonant regime. Three distinct qubit ensembles are seen.}%
\label{fig:5}%
\end{figure}

\section{Optical quantum metamaterials}

The domain of quantum metamaterials in the optical, or near IR, region of the spectrum is still in its infancy. As it has already been stated, some authors use the term quantum metamaterial to denote a structure in which quantum degrees of freedom are inserted \cite{Plumridge2007}. In some other cases, it is the expression "quantum dots metamaterials" that is used: this is to stress that, although quantum dots are inserted in a metamaterial, one is not interested in the quantum coherence of the dots, but rather on the gain that they provide, to counteract the losses due to the presence of metallic inclusions \cite{DCSEQDMM}. In other proposals, it is quantum wells that are inserted in a photonic structures. The quantum well are described electromagnetically by a permittivity allowing some control over the behavior of the structure. In ref. \cite{Plumridge2008a}, a layered metamaterial is investigated, in which the period comprises two GaAs quantum wells. This structure results in an effective permittivity tensor allowing to obtain a negative refraction. The effective properties strongly depend upon the 2D electron density in the quantum well. In ref. \cite{Plumridge2007} the same kind of structure is investigated in order to control plasmon propagation, allowing to obtain ultra-long propagation distances.

\begin{figure}%
\includegraphics[scale=1]{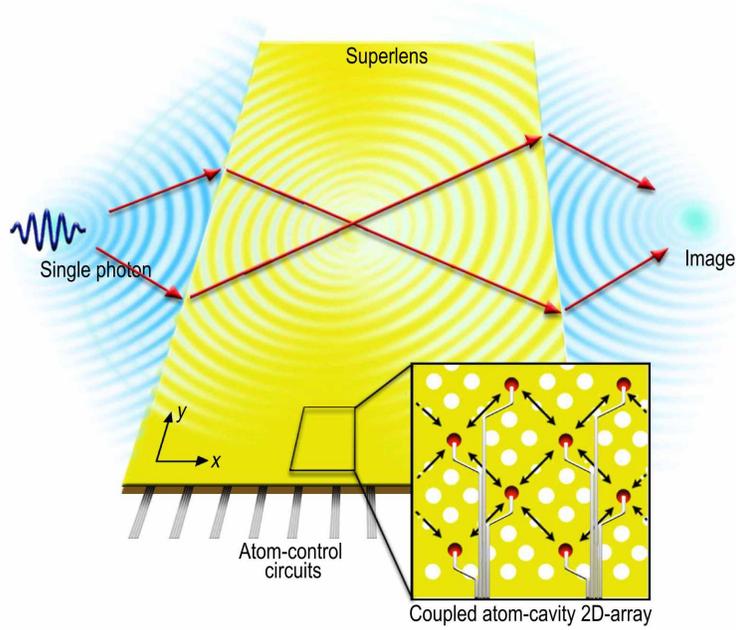}%
\caption{A cavity array metamaterial, taken from \cite{Quach2009}}%
\label{fig:6}%
\end{figure}

An original proposal was made in \cite{CPhthal} to extend the concept of metamaterial to quantum magnetism. The idea is to use molecular engineering or organic synthesis to fabricate magnetic quantum metamaterials. It is shown theoretically, by {\it ab initio} calculations, that CuCoPc2 (a chain of copper-phtalocyanine (CuPc) and cobalt phtalocyanine (CoPc)) possesses a relatively strong ferromagnetic interaction.

In the specific meaning used in this review, a proposal was made in (\cite{Quach2009,Quach2011}) to study the full quantum processes that occurs between the quantized electromagnetic field and two-level atoms. The system studied there is a 2D network of coupled atom-optical cavities, called a cavity array metamaterial (CAM). The authors propose to realize the model by using a two-dimensional photonic crystal membrane. The quantum oscillators could be quantum dots or substitution centers. 
Under reasonable assumptions, this system can be described by a Jaynes-Cummings-Hubbard Hamiltonian \cite{NRBH}. The system exhibits a quantum phase transitions \cite{QPTL, Quach2011} and it was proposed that it could be used as a quantum simulator. The effects of cloaking and negative refraction were also demonstrated. Quite naturally, the excitations of the system are hybrids of photonic and atomic states, namely polaritons. This kind of result is very well-known since the pioneering work of J. J. Hopfield \cite{hopfield}. Continuing along this way, an effective permittivity that is non-local in time and space could be derived, in exactly the same way as for natural material. This is the line followed in a series of papers by G. Weick where  a collection of metallic nanoparticles is shown to  exhibit collective plasmonic modes \cite{DPHLMP,TPPAIMP,DPBLMP}. However, a genuine quantum metamaterial requires more than that, namely the active coherent control of the quantum state of the "atoms" inserted in the photonic structure, so as to induce a control over the collective properties of the medium.

\begin{figure}
\begin{center}
\includegraphics[width=10cm]{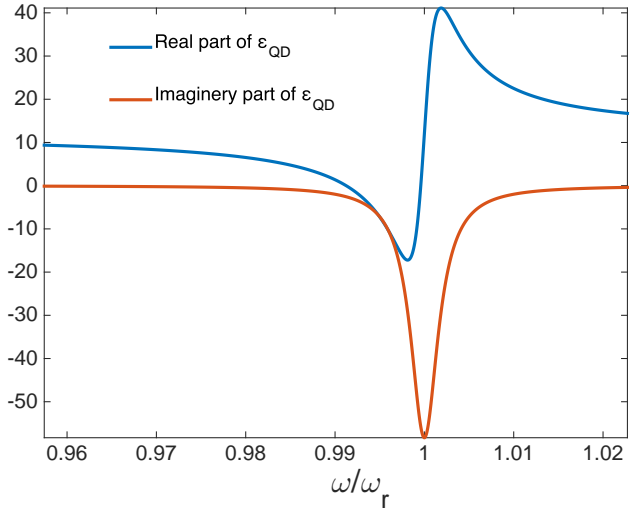}
\caption{The dielectric function of a core-shell quantum dot \cite{DFQDSCR}}
\label{fig:7}
\end{center}
\end{figure}

\begin{figure}
\includegraphics[width=\columnwidth]{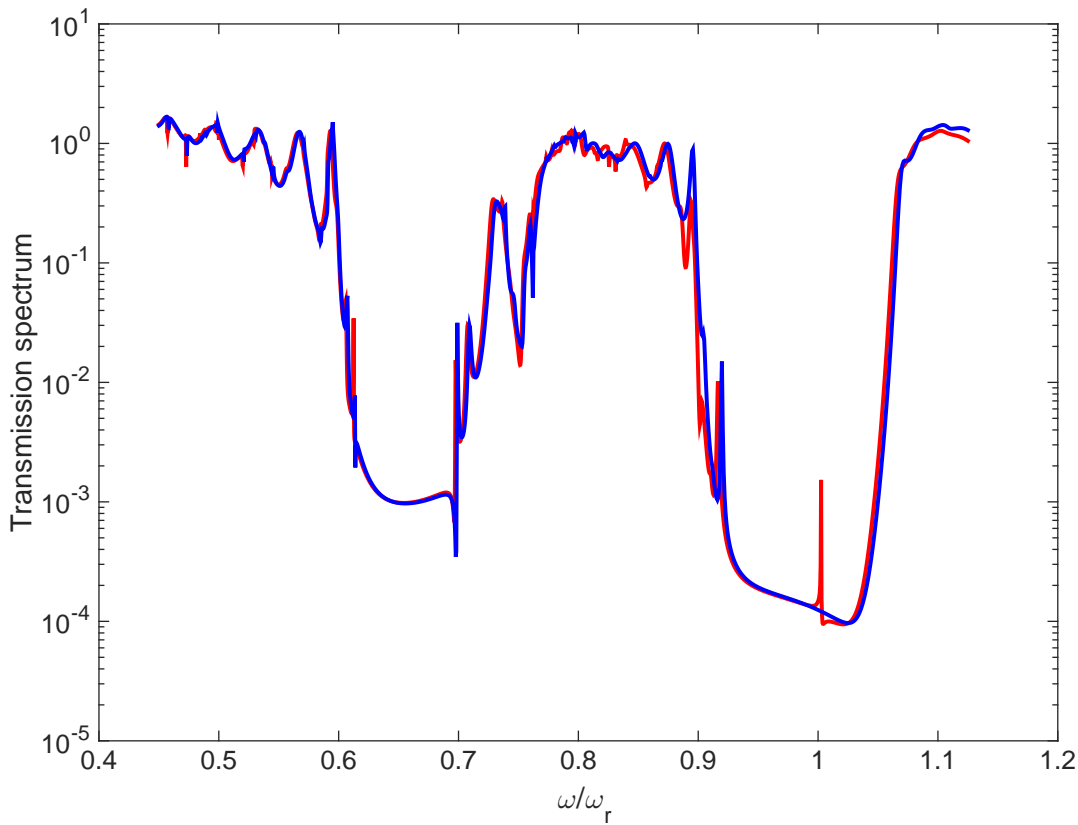}
\caption{Transmission spectrum the quantum metamaterial in the absorption regime (in blue) and in the emission regime (in red). The peak corresponds to the transition of the quantum dots.}
\label{fig:8}
\end{figure}

Such a system could be implemented by considering, as above, a photonic crystal in which quantum oscillators are inserted, under the guise of quantum dots for instance. The quantum dots can be described semi-classically by a dielectric function $\varepsilon_{QD}$ that reads as \cite{DFQDSCR}:
\begin{equation}
\varepsilon_{QD}(\omega)=\varepsilon_b+(f_c(E_e)-f_v(E_h)) \frac{a}{\omega^2-\omega_0^2+2i \omega \gamma}
\end{equation}
The prefactor $(f_c(E_e)-f_v(E_h))$ representing the difference between the populations of the levels can be either positive or negative. In the first case, it is in the absorption regime while in the latter it is in the emission (amplifying) regime.
The graph of $\varepsilon_{QD}(\omega)$ is given in fig. \ref{fig:7}.
The quantum dots can be grown inside dielectric nanopillars, and the nanopillars can be organized into a 2D periodic array, resulting into a photonic crystal with quantum dots. The bare photonic crystal is then tuned in such a way as to present a photonic band gap at the emission frequency of the quantum dots. The idea is then to realize a pump/probe experiment, where the pump controls the state of the quantum dots (absorption or emission). When the quantum dots are in the emission regime, a transmission peak appears in the transmission spectrum of the probe (Fig. \ref{fig:8}). This somewhat simplified model shows a macroscopic property, a conduction band, results from the quantum states of the microscopic quantum components. A full quantum treatment should be performed in order to address properly the quantum coherence of the system of the field coupled to the "atoms" of the metamaterial. As compared to the situations encountered in cavity quantum electrodynamics, the quantization of the electromagnetic field in open space comes with severe technical difficulties, if one follows the usual mode-decomposition path, abundantly described in all the literature devoted to field quantization. In fact, for open systems, Maxwell equations do not lead to hermitian eigenvalue problems and thus the eigenfunctions are not normalizable. A recent interesting approach is to use the so-called "quasi-normal modes" of the system \cite{lalanne}. 

\section{A review of the theoretical tools for quantum metamaterials in optics}

\begin{figure}
\includegraphics[width=12cm]{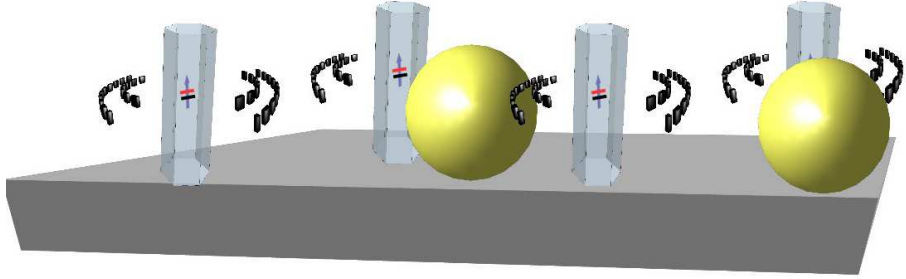}
\caption{A  schematic view of a possible quantum metamaterial in optics, made by several quantum dots embedded in semiconducting nanowires surrounded by metallic spheres. }
\label{fig:9}
\end{figure}

Here we review the tools of quantum optics for the description of the quantum dynamics of a collection of emitters (\textit{e.g.} quantum dots) in a complex electromagnetic environment that can be constituted by dielectric and/or metallic elements as depicted in Fig: \ref{fig:9}
 
For a single emitter in free space, the interaction with light is given by the minimal-coupling Hamiltonian that reads \cite{Milonni1984}:

$$ H = \frac{1}{2m}(\mathbf{p}-e\mathbf{A})^2 + eV(r) + \int d^3r' \mathbf{E}^{\bot 2}+\mathbf{B}^2 $$

where $\mathbf{p}$ is the electron momentum, $\mathbf{A}$ the vector-potential, $V(r)$ the binding potential for the electron, $\mathbf{B}$ the magnetic field and $\mathbf{E}^{\bot}$ the transverse part of the electric field, satisfying $\mathbf{\nabla . E^{\bot}} = 0$ (c.f. \cite{Jackson} p. 254). The emitters are coupled to each other through the electromagnetic field that comprises both radiating and evanescent terms. 

For neutral emitters in a complex electromagnetic environment, one may need a mesoscopic description, where the emitters are described through their multipole moments instead of the dynamics of the electric charges. Moreover, it can be convenient to describe the electromagnetic environment by polarization and magnetization fields instead of charge and current densities \cite{Jenkins}. Last but not least, a Hamiltonian involving the physical fields $\mathbf{E}$ and $\mathbf{B}$ rather than the potentials $\varphi(\vec{r})$ and $\mathbf{A}(\vec{r})$ can also be more convenient. Applying the Power-Zienau-Wooley transform to the previous Hamiltonian\cite{Power, Milonni1984, Jenkins} (see also \cite{Cohen} p. 282) fulfills these requirements.  The Power-Zienau-Wooley transform leads to the following Hamiltonian(see Milonni\cite{Milonni} p.121):

\begin{center}
\begin{eqnarray}
H = H_{mat} + H_{field} + H_{int} 
\label{Eq:HPZ}
\end{eqnarray}
\end{center}

where
\begin{itemize}
\item $ H_{mat} = \frac{\mathbf{p}}{2m} + V(r) + \int d \vec{r} ~ \frac{\mathbf{P}^{\bot 2}}{2\varepsilon_0}  $ is the Hamiltonian describing the dynamics of the atom variables, $\mathbf{P}$ is the polarization field. Even if the term $\int d \vec{r} ~\mathbf{P}^{\bot 2}/2\varepsilon_0$  is important to reproduce the correct dynamics of the emitter\cite{Milonni1984}, it is usually neglected when studying the interaction between the emitter and light.  It is usually argued that this term merely shifts the energy levels, an effect that can be accounted for by a correct renormalization of the emitter energy levels. Nevertheless in the ultra-strong coupling regime, this term has to be taken into account \cite{DeLiberato2014}. It leads to a decoupling of matter and light states because of a screening of the incident light by the polarization field $\mathbf{P}$, resulting for example in a reduction of the Purcell factor, while increasing the coupling between the field and the emitter \cite{DeLiberato2014}. 

\item $ H_{field} = \int  \frac{ \mathbf{D}^{2}(\vec{r})}{2\varepsilon_0}  + \frac{\mathbf{B}^2(\vec{r})}{2\mu_0}$ is the Hamiltonian describing the  electromagnetic field dynamics. 
 
\item $ H_{int} = -\int  \mathbf{P}(\vec{r}).\mathbf{D}(\vec{r})~ d\vec{r} $ is the Hamiltonian describing the interaction between light and matter.  
\end{itemize}

$\mathbf{D}(\vec{r})$ is the displacement vector. Note that the Hamiltonian given by the equation eq.(\ref{Eq:HPZ}) is exact. It is completely equivalent to the minimal-coupling hamiltonian (Cohen\cite{Cohen} p.298). From Hamiltonian eq.(\ref{Eq:HPZ}), with the help of the Heisenberg equation, one can find the dynamical equations satisfied by each operators (field and matter operators). 

Concerning the field operators, it is assumed that they are related to each other through Maxwell equations. This assumption implies that there exists some commutators between the field operators (p.18 in\cite{Leonhardt}): $[\mathbf{D}(\vec{r},t),\mathbf{A}(\vec{r'},t)] = \frac{i \hbar}{\varepsilon_0} ~ \delta^{T}(\vec{r}-\vec{r'})$ where $\delta^{T}(\vec{r})$ is the transverse delta distribution (See also \cite{Cohen} p. 233). Finally, one gets the following set of equations between the field operators:
\begin{equation*}
\begin{array}{l}
\nabla \times \mathbf{E} = - \partial_t \mathbf{B}\\
\nabla \times \mathbf{B} = \mu_0 \partial_t \mathbf{P}+ \frac{1}{c^2} \partial^2_{t^2} \mathbf{E}
\end{array}
\end{equation*}
One arrives at the well-known wave equations satisfied by the electric-field operator:

\begin{equation}
\nabla \times \nabla \times \mathbf{E} - \frac{1}{c^2} \frac{\partial^2 \mathbf{E}}{\partial t^2}= - \mu_0 \frac{\partial^2 \mathbf{P}}{\partial t^2} 
\label{Eq:WaveEq}
\end{equation}

Concerning the dynamics of the matter degrees of freedom, some usual approximations are done. We write the polarization field as a sum of a polarization field due to the atoms $\mathbf{P}_a$ and a polarization field due to the electromagnetic environment $\mathbf{P}_{ind}$: $\mathbf{P} = \mathbf{P}_a + \mathbf{P}_{ind}$.  We assume that the
polarization field due to the electromagnetic environment $\mathbf{P}_{ind}$ responds linearly and locally to the electric field. It reads as\cite{Vogel} 
$$\mathbf{P}_{ind}(\vec{r},t) = \int_{-\infty}^{+\infty} \varepsilon_0 \chi(\vec{r},t')\mathbf{E}(\vec{r},t')dt'$$
where $\chi(\vec{r},t')$ is the susceptibility at position $\vec{r}$ and time $t$.

Concerning the polarization field due to emitters, we work in the usual dipole approximation and approximate the polarization field by keeping only the first term in the multipolar expansion even if this approximation can be crude for quantum dots\cite{Yan2008}. If there are $N_{\alpha}$ emitters, the polarization field due to the emitters is then written as $\mathbf{P}_{a}(\vec{r}) = \sum_{\alpha=1}^{N_{\alpha}}  \mathbf{d}^{\alpha} \delta(\vec{r} - \vec{r_{\alpha}})$ where $ \mathbf{d}^{\alpha}$ and $\vec{r_{\alpha}}$ are respectively the dipole-moment operator and the position of the emitter labeled by $\alpha$, and $\delta(\vec{r})$ is the usual Dirac distribution.
It is convenient to express all matter-operators with the help of the basis defined by the eigenstates of the matter hamiltonian $H_{mat}$. The matter hamiltonian $H_{mat}$ is written as $H_{mat} = \sum_{\alpha=1}^{N_{\alpha}} H_{mat}^{\alpha}$  where $H_{mat}^{\alpha}$ is the hamiltonian of the $\alpha^{th}$ emitter. We note $\{\ket{\alpha,i}\}$ the eigenbasis constructed from the eigenstates of $H_{mat}^{\alpha}$ that satisfied $H_{mat}\ket{\alpha,i} = E^{\alpha}_i \ket{i}$ and the completeness condition $\sum_{i} \ket{\alpha, i}\bra{\alpha,i} = I_d^{\alpha}$, where $I_d^{\alpha}$ is the identity matrix acting on the subspace of the $\alpha^{th}$ emitter. 

We now assume that emitters are two-level systems. The ground state is labelled by $i=-$ whereas the excited level is labelled by $i=+$. Applying twice the completeness condition on the hamiltonian $H_{mat}^{\alpha}$, one finds $H_{mat}^{\alpha} = \frac{E_{-}+E_{+}}{2}I_d^{\alpha} + \frac{\hbar \omega^{\alpha}}{2} \sigma^{\alpha}_{z} $ where $\hbar \omega^{\alpha} = E_{+}^{\alpha} - E_{-}^{\alpha}$ and $\sigma_z^{\alpha} = \ket{\alpha,+}\bra{\alpha,+} - \ket{\alpha,-}\bra{\alpha,-}$. $\sigma_z^{\alpha}$ acts on the subspace defined by the eigenvectors of the $\alpha^{th}$ emitter and measured its population difference between the excited and the ground state. The first term in $H_{mat}^{\alpha}$ is a constant that can be omitted by choosing correctly the reference of the energy. We then write the matter hamiltonian as (see\cite{Carmichael} p. 23 and \cite{Milonni} p.128): $$H_{mat} = \sum_{\alpha = 1} ^{N_{\alpha}} \frac{\hbar \omega^{\alpha}}{2} \sigma^{\alpha}_{z}$$ .

The polarization field due to the emitters can also be written with the help of the eigenbasis of each emitter by writing the dipole-moment operator in this basis: 
$$ \mathbf{d}^{\alpha} =  \vec{d}^{\alpha}_{+-} \ket{\alpha,+}  \bra{\alpha,-} + \vec{d}^{\alpha}_{-+} \ket{\alpha,-}  \bra{\alpha,+} $$

where $\vec{d}^{\alpha}_{+-} = \bra{\alpha,+} \mathbf{d}^{\alpha}\ket{\alpha,+}$ is the projection of the dipole-moment operator in the basis $\{ \ket{\alpha,+}, \ket{\alpha,-} \}$. The diagonal elements are null because we assume that the emitters have no permanent dipole.
We introduce the raising $\sigma^{\alpha}_+ = \ket{\alpha,+}\bra{\alpha,-}$ and lowering operators $\sigma^{\alpha}_- = \ket{\alpha,-}\bra{\alpha,+}$. Finally, the polarization field due to the atoms reads:
$$ \mathbf{P}_{a}(\vec{r}) = \sum_{\alpha=1}^{N_{\alpha}} (\vec{d}^{\alpha}_{+-} \sigma^{\alpha}_{+} + \vec{d}^{\alpha}_{-+} \sigma^{\alpha}_{-})   \delta(\vec{r} -  \vec{r_{\alpha}}) $$

If we assume that the dipole moment projection $\vec{d}^{\alpha}_{+-} = (\vec{d}^{\alpha}_{-+})^\star = \vec{d}_0\alpha$ is real, the polarization field simplifies with the help of the Pauli matrice $\sigma_x$:

$$ \mathbf{P}_{a}(\vec{r}) = \sum_{\alpha=1}^{N_{\alpha}} \vec{d}^{\alpha}_{0} \sigma^{\alpha}_{x} \delta(\vec{r} -  \vec{r_{\alpha}}) $$

since $\sigma_x = \frac{1}{2}(\sigma^{\alpha}_{+} + \sigma^{\alpha}_{-})$ (see \cite{Carmichael} p.24, \cite{Allen} p.35).

With the help of a Fourier transform, the hamiltonian for the free electromagnetic field can be written $H_{field} = \sum_{j,s} \hbar\omega_j b_{j,s}^+b_{j,s}$ where $b_j,s$ (\textit{resp.} $b^+_{j,s}$) is the annihilation (\textit{resp.} creation ) operator of the electromagnetic mode $j$ with polarization $s$. Within this decomposition the electric field on its own reads  $\mathbf{E} = \sum_{j,s} e_{j,s}(b_{j,s} - b^+_{j,s})$. Finally, following these successive approximations one finds the hamiltonian given by the equation eq.(\ref{eq:1})

The behavior of the quantum metamaterial can be computed by solving simultaneously the equation eq.(\ref{Eq:WaveEq}) and the equations of motions for the collection of two-level "atoms" given by \cite{Milonni1984}, p.37 and \cite{Allen} :

\begin{center}
\begin{eqnarray}
\nonumber
 \dot{\sigma_x^\alpha} &=& -\omega^\alpha \sigma_y^{\alpha} \\
\nonumber
\dot{\sigma_y^\alpha} &=& \omega^\alpha \sigma_x^{\alpha} + \frac{2}{\hbar} \vec{d}_0^\alpha.\mathbf{E}(\vec{r}_{\alpha}) \sigma_z^{\alpha}  \\
\nonumber
\dot{\sigma_z^\alpha} &=& - \frac{2}{\hbar} \vec{d}_0^\alpha.\mathbf{E}(\vec{r}_{\alpha}) \sigma_y^{\alpha} 
\end{eqnarray}
\end{center}
  
Where all terms proportional to $\mathbf{P}^2(\vec{r})$ have been neglected. These equations are non-linear coupled differential equations since $\sigma_x^\alpha$ are sources of the electric field eq:(\ref{Eq:WaveEq}). The electromagnetic environment contributes also as a source term in eq:(\ref{Eq:WaveEq}). This source term is important since it is responsible for all the unusual effects demonstrated theoretically or experimentally with classical metamaterials. Its effect on quantum metamaterials has been barely studied since solving this system of equations is challenging, even for a very simple geometry\cite{Braak2011}. New theoretical tools should be developed to accurately describe the behavior of a quantum metamaterial in a complex electromagnetic environment.  Nevertheless more approximations can be done to solve these equations. One can use the single electromagnetic-mode approximation\cite{Quach2009,Quach2011,Quach2013,Everitt2014} or perform a semi-classical approximation\cite{Rakhmanov2008,Zagoskin2009,Zagoskin2011,Asai2015,McEnery2014}. In the latter, it is assumed that there are no correlations \cite{Allen} between the electromagnetic field and the matter degrees of freedom.

\section{Conclusions} 

The field of quantum metamaterials research arose at the intersection of quantum optics, microwave and Josephson physics, and quantum information processing. One of its rather paradoxical feature is that, while the theoretical progress in this area still significantly outweighs the experiment, the theoretical challenges seem more significant. Indeed, the existing experimental techniques, especially in case of superconducting structures, already allow creating massive arrays. The 20-qubits prototype \cite{Macha2014} is much smaller than a recently fabricated 1000+-qubits superconducting quantum annealer D-Wave 2X. Given a simpler structure, and less strict demands to a quantum metamaterial than to a quantum computer, making and testing quantum metamaterials on this scale is a question of time and funding. On the other hand, the theoretical analysis of quantum metamaterials produces promising results, already using simple approximations. Nevertheless the understanding of the full scale of effects which can be expected in these systems requires a more detailed analysis of large scale quantum coherences and entanglement. Because of the well-known impossibility to effectively simulate a large quantum system by classical means, a direct approach to this is currently limited to structures containing (optimistically) less than a hundred qubits. New theoretical tools need to be developed, generalizing the methods of quantum theory of solid state \cite{Zagoskin2013a}.  

These challenges also present alluring opportunities. Developing and testing new theoretical methods applicable to large quantum coherent systems would be valuable for the whole field of quantum technologies, including quantum computing. Optical elements based on quantum metamaterials would provide new methods for image acquisition and processing. Last but not least, a quantum metamaterial would be a natural test bed for the investigation of quantum--classical transition, which makes this class of structures interesting also from the fundamental point of view.

\acknowledgements

AZ was supported in part by  the EPSRC grant EP/M006581/1 and by the Ministry of Education and Science of the Russian Federation in the framework of Increase Competitiveness Program of NUST«MISiS»(No. K2-2014-015). 

\bibliography{QMM-REVIEW-resub}


\end{document}